\documentclass[twocolumn,aps,prb,floats,showpacs,superscriptaddress]{revtex4}
\usepackage{amsmath}

\newcommand{\be}{\begin{equation}}
\newcommand{\ee}{\end{equation}}

\usepackage{epsfig}
\usepackage{graphicx}
\usepackage{bm}
\usepackage{amssymb}%

\def\gapp{\lower.35em\hbox{$\stackrel{\textstyle>}{\sim}$}}
\def\lapp{\lower.35em\hbox{$\stackrel{\textstyle<}{\sim}$}}

\begin{document}

\bibliographystyle{apsrev}

\title{Large conduction band and Fermi velocity spin splittings due to Coulomb interactions in single-layer MoS$_{2}$}
\author{Yago Ferreiros}
\email{yago.ferreiros@csic.es}
\author{Alberto Cortijo}
\email{alberto.cortijo@csic.es}

\affiliation{Instituto de Ciencia de Materiales de Madrid, CSIC, Cantoblanco, 28049 Madrid, Spain.}

\begin{abstract}
We study the effect of Coulomb interactions on the low energy band structure of single-layer transition metal dichalcogenide semiconductors using an effective low energy model. We show how a large conduction band spin splitting and a spin dependent Fermi velocity are generated in MoS$_2$, as a consequence of the difference between the gaps of the two spin projections induced by the spin-orbit interaction. The conduction band and Fermi velocity spin splittings found are in agreement with the optical absorption energies of the excitonic peaks A, B measured in the experiments.
\end{abstract}
\pacs{31.70.-f, 73.22.-f, 78.67.-n}

\maketitle

\section{Introduction}

In the post-graphene era\cite{CGP09}, the search of new low dimensional materials has placed the transition metal dichalcogenides (TMDCs) in a prominent situation. Known for decades, these materials share attracting features common to graphene: they are layered materials with strong covalent bonding within layer and weak Van der Waals interlayer forces, being perhaps the most known member the molybdenum disulphide (MoS$_{2}$)\cite{MLH10}. A subset of the large family of TMDCs are semiconductors, with sizeable direct gaps ranging from one to several eV around the K and K' points of the Brillouin zone (BZ)\cite{XLF12}. The presence of a gap in the band structure of these systems is a feature that distinguish them from graphene and makes these materials highly valuable for electronic and optoelectronic applications. 

Apart from their potential applicability in electronics, TMDCs monolayers are also an attractive arena of research in the field of spintronics. The transition metals forming the TMDCs display a rather large intra-atomic spin-orbit interaction. Together with the absence of inversion symmetry of the crystalline structure of TMDCs monolayers, this induces a spin-splitting between the two (otherwise degenerate) spin projections in the band structure\cite{XLF12}. Due to time reversal symmetry requirements, this spin splitting is opposite in both valleys and consequently it allows for controlling valley population employing circularly polarized light\cite{MHK12}. 

The orbital nature of the electronic states around the K, K' points indicates that the effect of the spin-orbit interaction is quite different for the conduction band (formed predominantly by the $d_{3z^{2}-r^{2}}$ orbital, with $m_{l}=0$) and the valence band (mostly made of a linear combination of the $d_{x^{2}-y^{2}}$ and $d_{xy}$ orbitals, with $m_{l}=\pm 2$). Such a particular atomic band population implies that the splitting of the valence band is first order in the spin-orbit interaction, while the splitting of the conduction band is second order in the mentioned orbitals and a very small contribution of first order processes of higher energy orbitals\cite{OR13,KZD13}. According to several Density Functional Theory (DFT) calculations\cite{KGF13,XLF12,KZD13}, this favors a weak spin splitting in the conduction band, of the order of a few meV, and a considerably larger splitting in the valence band, of the order of hundreds of meV. On the experimental side, characterization of the low energy band structure of TMDC semiconductors is still incomplete. The most common procedure to determine the parameters entering in the band structure description, that are later employed in other methods (e. g. in tight binding calculations\cite{CRS13,RMA13}), consists in contrasting experimental data obtained by optical means with theoretical results obtained by solving the GW-corrected Kohn-Sham equations in its several variations with different degrees of success\cite{QJL13}. 

Here we follow an alternative route: we will use a self consistent GW treatment of the Coulomb interaction together with the effective Hamiltonian around the K, K' points, using experimental data and physically motivated considerations to determine the band structure parameters of the system. We will find, for MoS$_2$, that a self consistent treatment of the many-body problem based on an unscreened Coulomb interaction, together with an effective low energy model for the electrons around the $K$ and $K'$ points, give much larger values for the conduction spin splitting $\lambda_{c}$ than the ones reported by \emph{ab initio} calculations, and, importantly, these values are fully consistent with the experimental values reported for the excitonic spectrum.

Our results are pertinent since recently a realization of the Quantum Spin Hall (QSH) phase has been proposed using TMDCs as a viable platform\cite{COG14}. In this proposal the system has to be doped with holes in order to take into account the large value of the spin splitting in the valence band. However, doping with holes turns out to be much more difficult than doping with electrons from an experimental perspective. Our findings suggest that it is possible to realize such phase also doping with electrons since our results for $\lambda_{c}$ are considerably larger than the previously reported in the literature.

The paper is organized as follows: In Sec.\ref{sec:model} we describe the effective continuum model that captures the essential low energy physics of TMDCs together with the Coulomb interaction. In Sec.\ref{sec:SDeqs} we define the elements for the self-consistent treatment of the Schwinger-Dyson equation and the quantum corrections to be obtained. In Sec.\ref{sec:expdata} we give a physical description of the conditions that we will employ to reduce the number of free parameters to just two: the coupling constant and the momentum cut-off, putting special emphasis in which conditions come from experimental measurements and which ones have to be fixed by some other physical insight. In Sec.\ref{sec:solution} we discuss the most salient qualitative features of the solutions obtained, finishing with Sec.\ref{sec:summary}, where we give a brief account of the results obtained.

\section{The model}
\label{sec:model}

We start from the following low energy bare Hamiltonian density for single-layer TMDC semiconductors close to the K, K' points\cite{XLF12}:
\begin{eqnarray}
\mathcal{H}_0&=&\psi^{+}\left((\tau\sigma_xp_x+\sigma_yp_y) V^{0}_\tau+\frac{\Delta^{0}}{2}\sigma_z s_{0}\right)\psi +\nonumber\\
&+&\psi^{+}\left(\frac{\lambda^{0}_c}{2}\tau(\sigma_{0}+\sigma_z)s_z+\frac{\lambda^{0}_v}{2}\tau(\sigma_{0}-\sigma_z)s_z\right)\psi,\label{initialHam}
\end{eqnarray}
where $(s_{0}\equiv \mathbf{1}$, $s_{x,y,z})$ and $(\sigma_{0}\equiv \mathbf{1}$, $\sigma_{x,y,z})$ are the Pauli and identity matrices for the spin and the sublattice degrees of freedom, respectively, $2\lambda^{0}_{c}$ ($2\lambda^{0}_{v}$) is the conduction (valence) band splitting, $\tau=\pm1$ is the valley index, and the matrix $V^{0}_\tau$ is:
\begin{equation}
V^{0}_{+1}=\left(\begin{array}{cc}v^{0}_+&0\\0&v^{0}_-\end{array}\right);\quad V^{0}_{-1}=\left(\begin{array}{cc}v^{0}_-&0\\0&v^{0}_+\end{array}\right)
\end{equation}
with $v^{0}_\pm$ the Fermi velocities for spin up/down $(+/-)$ electrons. The superscript $0$ is the notation used for the bare parameters. Although the Fermi velocities for both spins are assumed to be equal in the absence of interactions (since the hopping parameters should be insensitive to spin), we define the bare Fermi velocity for each spin separately, because as we will see, quantum corrections renormalize each velocity differently. Also we neglect terms of second order in momentum in a $\mathbf{k}\cdot\mathbf{p}$ scheme\cite{RMA13} that are not relevant in the discussion in the next sections.

Coulomb interaction will be modeled by a coupling to an auxiliary scalar field $\varphi$\cite{GGV99}:
\begin{equation}
\mathcal{H}_{int}=e\psi^\dagger\psi \varphi+\epsilon\varphi|\vec{\nabla}|\varphi,\label{Coulombfield}
\end{equation}
where $\epsilon$ is the dielectric permitivity (we use units $\hbar=1$ at the intermediate stages, restoring standard units at the end of the calculations). 

\section{Schwinger-Dyson equations}
\label{sec:SDeqs} 

To self-consistently  find the effects of the interaction on the low energy band structure we shall make use of the Schwinger-Dyson equation for the electron propagator\cite{D49,S51,ABK86}:
\begin{equation}
\Sigma_\tau(p)=e^2\int\frac{d^3q}{(2\pi)^3}D(q)G_\tau(p-q),
\label{SD equation}
\end{equation}
where $D(q)^{-1}=\epsilon|\vec{q}|+\Pi(q)$ is the (inverse of the) dynamically screened Coulomb interaction (with $\Pi(q)$ the dressed polarization function) and $G_\tau(p)^{-1}=G_0^{-1}(p)+\Sigma_\tau(p)$ is the (inverse of the) full fermion propagator. The form of the self-energy $\Sigma_\tau(p)$ can be parametrized in terms of corrections to the bare parameters defined in Eq.(\ref{initialHam}): 
\begin{eqnarray}
\Sigma_\tau(p)&=&-(\tau\sigma_xp_x+\sigma_yp_y)\delta V_\tau-\frac{\delta\Delta}{2}\sigma_zs_{0}-\nonumber\\
&-&\frac{\delta\lambda_c}{2}\tau(\sigma_{0}+\sigma_z)s_z-\frac{\delta\lambda_v}{2}\tau(\sigma_{0}-\sigma_z)s_z.
\end{eqnarray}
The renormalized parameters are defined as the sum of the bare parameters plus the quantum corrections:
\begin{eqnarray}
\Delta=\Delta^{0}+\delta\Delta , \quad\lambda_{c,v}=\lambda_{c,v}^{0}+\delta\lambda_{c,v}, \quad v_\pm =v_\pm^{0}+\delta v_\pm.
\end{eqnarray}
Only terms up to order one in momentum have been taken into account in the definition of the self-energy, being consistent with the low energy expansion of the effective Hamiltonian (\ref{initialHam}), which is also first order. Note also that in eq. (\ref{SD equation}) we could have chosen a different run of momentum $D(p-q)G_\tau(q)$, as both runs are consistent with momentum conservation, and both give the same final result.

The computation of $\Pi(q)$ to leading order in a $1/N$ expansion, where N is the number of fermion flavors ($N=2$ in our case, one for each valley), is a standard calculation and the result reads\cite{R84,ABK86}:
\begin{eqnarray}
\Pi(q)=\frac{e^2}{4\pi}|\vec{q}|^2\sum_{s=\pm}\left[ \frac{2m_s}{q^{2}_s}+\frac{q^2_s-4m_s^2}{q^{3}_{s}}\arctan\left(\frac{q_s}{2m_s}\right)\right],\label{polarizationfunction}
\end{eqnarray}
where $s=\pm$ denotes the spin degrees of freedom, $q^2_s=q_0^2+v^2_s|\vec{q}|^2$, $m_+=(\Delta+\lambda_c-\lambda_v)/2$ and $m_-=(\Delta-\lambda_c+\lambda_v)/2$. 

In what follows we will work with only one of the two valleys: $\tau=+1$, since it can be easily seen that the equations obtained for one valley are equivalent to those obtained for the other valley just by changing the sign of the spin splittings. The next step is to compute:
\begin{equation}
e^2\int\frac{d^3q}{(2\pi)^3}D(q)G_{+1}(p-q)\equiv\left(\begin{array}{cc}I_+(p)&0\\0&I_-(p)\end{array}\right),\label{integral}
\end{equation}
with:
\begin{equation}
I_s(p)=-I^{z}_s\sigma_z-v_s I_s(\sigma_xp_x+\sigma_yp_y)+\mathcal{O}(p^2)\label{Ifunctions}
\end{equation}
Performing the integral in Eq.(\ref{integral}) we obtain (see Appendix \ref{app} for details on the computation of the fermionic self-energy):
\begin{subequations}
\begin{equation}
I^{z}_s=\frac{3m_r}{4}\ln\left(1+\frac{2g_sm_s}{3m_r}\right)+\frac{g_sm_s}{2+\pi g_r}\ln\left(\frac{\Lambda v_s}{m_s}\right),
\end{equation}
\begin{equation}
I_s=\frac{3m_r}{4m_s}\ln\left(1+\frac{2g_sm_s}{3m_r}\right)+\frac{g_s}{4+2\pi g_r}\ln\left(\frac{\Lambda v_s}{m_s}\right),
\label{eq. Fermi velocity}
\end{equation}
\end{subequations}
where $\Lambda$ is a momentum cut-off and we have defined:
\begin{equation}
g_s=e^2/(4\pi\epsilon v_s);\quad g_r=e^2/(4\pi\epsilon v_r),\nonumber
\end{equation}
\begin{equation}
m_r=\frac{2\, m_+ m_-}{m_++m_-};\quad v_r=\frac{2\, v_+ v_-}{v_++v_-}.
\end{equation}

\begin{figure*}
(a)
\begin{minipage}{.46\linewidth}
\includegraphics[scale=0.36]{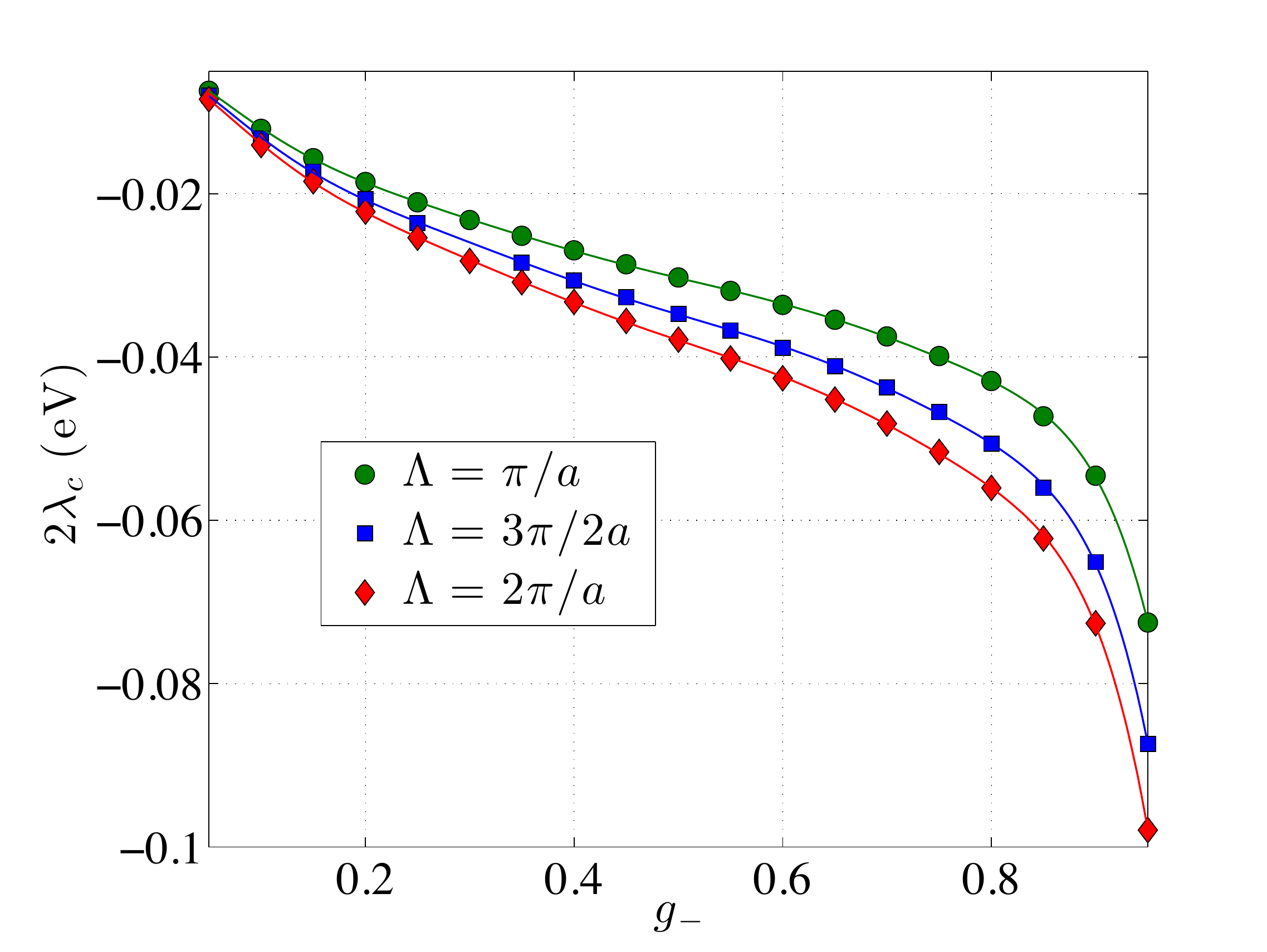}
\end{minipage}
(b)
\begin{minipage}{.46\linewidth}
\includegraphics[scale=0.36]{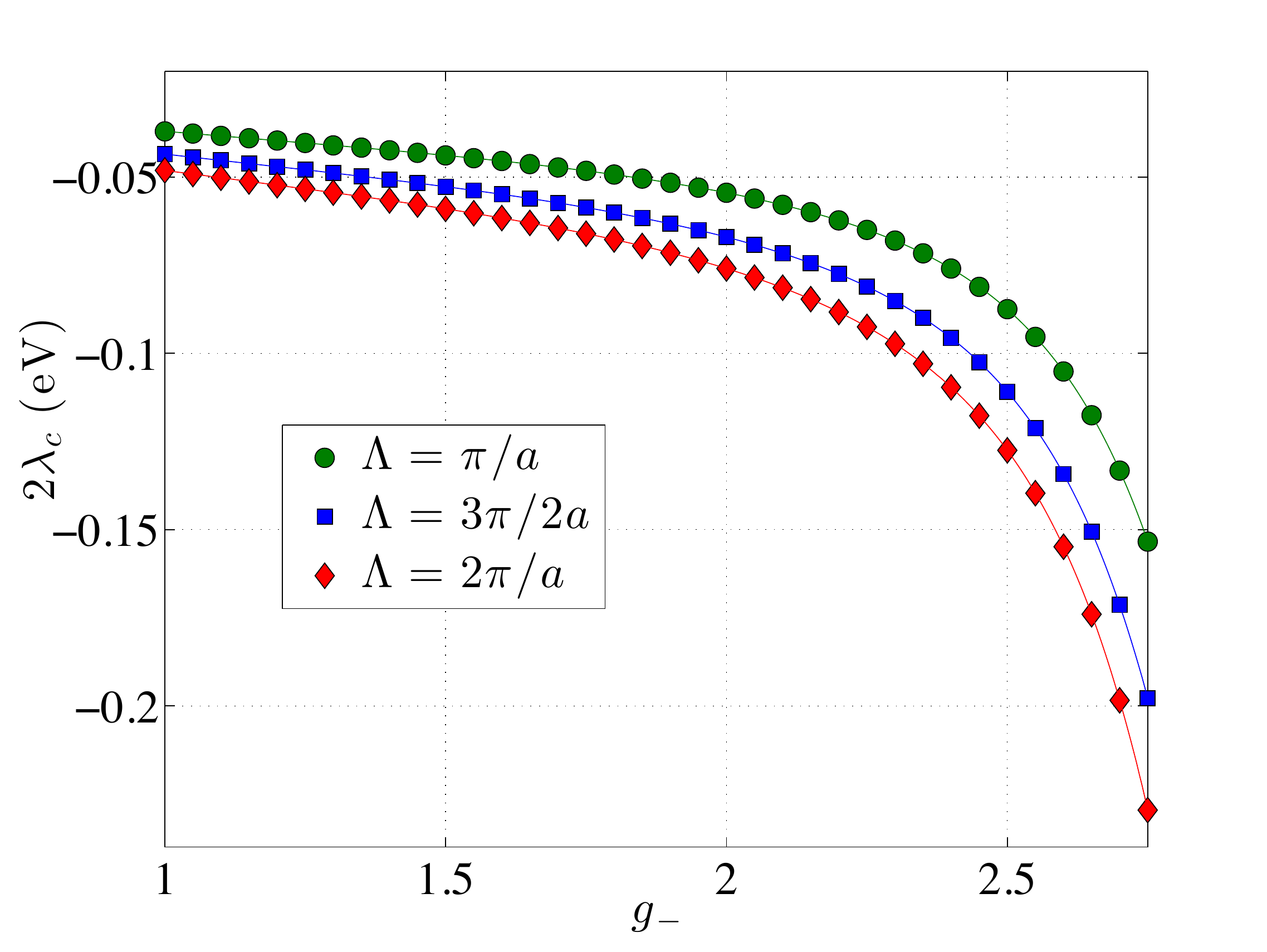}
\end{minipage}
(c)
\begin{minipage}{.46\linewidth}
\includegraphics[scale=0.36]{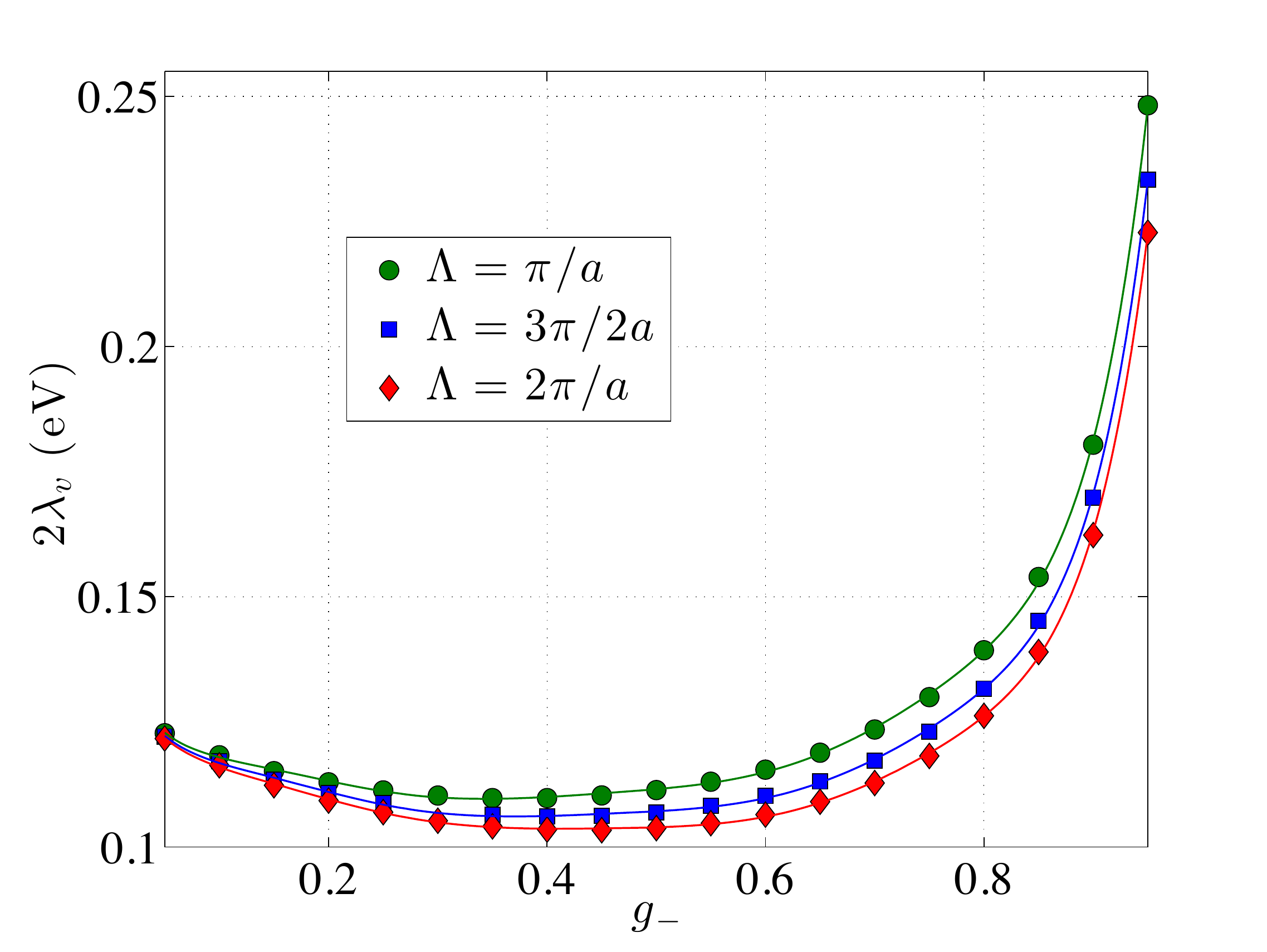}
\end{minipage}
(d)
\begin{minipage}{.46\linewidth}
\includegraphics[scale=0.36]{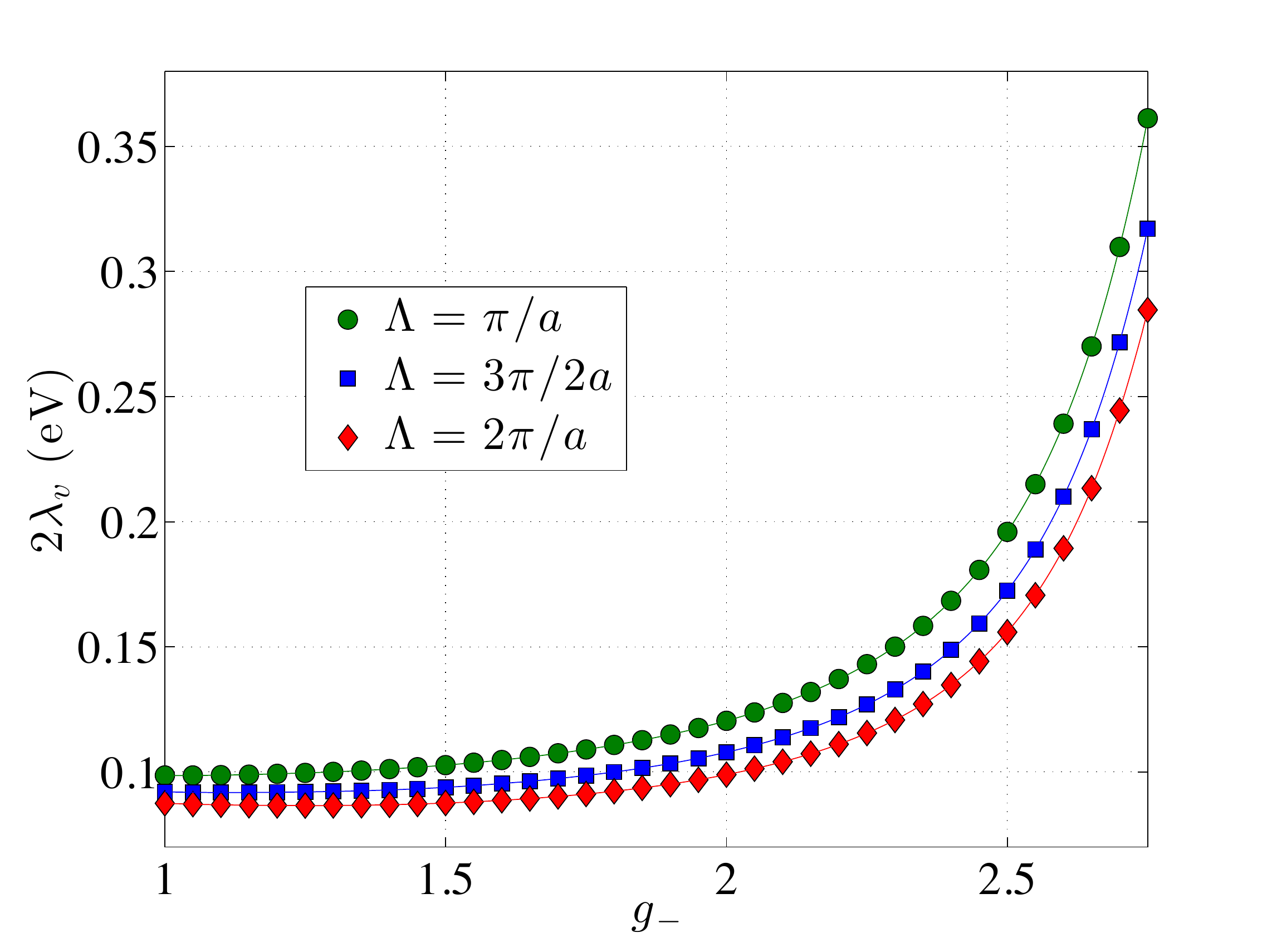}
\end{minipage}

\caption{Values for the $\tau=+1$ valley of the absolute value of conduction band splitting (a,b) and valence band splitting (c,d) as a function of the corrected coupling constant for spin down electrons $g_-$, for three different values of the momentum cut-off (in units of $\hbar$).}
\label{fig. 1}
\end{figure*}

In terms of the quantities $I^{z}_{s}(p)$, and $I_{s}(p)$, the Schwinger-Dyson equations are written as:
\begin{equation}
\Sigma(p)=\left(\begin{array}{cc}I_+(p)&0\\0&I_-(p)\end{array}\right)\label{SDeqs}.
\end{equation}
Eq. (\ref{SDeqs}) is a set of four equations for the diagonal elements (two for each spin), from which only three are linearly independent, and four equations for the off-diagonal elements, from which only two (one for each spin) are linearly independent. Thus we have five linearly independent equations for five variables:
\begin{eqnarray}
\frac{\delta\Delta}{2}+s\delta\lambda_c=I^{z}_s &,&  \frac{\delta\Delta}{2}-\delta\lambda_v=I^{z}_{+}, \nonumber\\
\delta v_s&=&v_sI_s.
\label{SDeqs2}
\end{eqnarray}
From these equations we can obtain the quantum corrections $\delta\Delta,\delta\lambda_{c,v},\delta v_\pm$ as functions of the bare parameters $\Delta^{0}, \lambda^{0}_{c,v}, v^{0}_{\pm}$, and the cut-off $\Lambda$.

\section{Matching the experimental data}
\label{sec:expdata}
To eliminate the dependence of the quantum corrections on the bare parameters and the cut-off one needs to impose renormalization conditions, which should be obtained from experiments. In the case of single layer TMDC semiconductors one of the most relevant data are the energies of two excitonic peaks A, B found in absorbance experiments in MoS$_2$\cite{MLH10,SSZ10,MHL13,MHK12}, which provide two renormalization conditions. 

We will focus our attention on MoS$_2$, as this is the most studied compound of the family of TMDCs, both theoretically and experimentally. For MoS$_2$ on a quartz substrate, the excitonic peaks A, B found in absorbance experiments are centered at energies of $E_A=1.85$ eV and $E_B=1.98$ eV\cite{SSZ10}. The absorption energy difference between these two peaks is a direct consequence of the splitting of the spin up and spin down bands. \emph{Ab initio} calculations give a very small conduction band splitting of approximately $2|\lambda_c|\sim3$ meV \cite{KH12,KZD13,KGF13,KF13,QJL13,CL12,MSH13}. Hence, according to first principles-based calculations, the difference between the optical absorption energies of the peaks A and B is related almost entirely to a large value of the valence band splitting\cite{QJL13,CL12,MSH13}. Even analytical calculations of the excitonic properties of MoS$_2$ use as an input the parameter values obtained by \emph{ab initio} calculations\cite{BM13}.

The optical absorption energies of the excitonic bound states are obtained by solving the two particle problem for the Dirac equation. The expression for the optical absorption energies of the excitons is\cite{KH98,RC13}:
\begin{equation}
E_s=m_s\left(1+\frac{n+\sqrt{j^2-g_s^2/4}}{\sqrt{g_s^2/4+\big(n+\sqrt{j^2-g_s^2/4}\big)^2}}\right)
\label{eq exc energies}
\end{equation}
where $n=0,1,2...$ is the principal quantum number and $j=\pm1/2,\pm3/2...$ is the angular quantum number. The two lowest energy excitonic configurations are $n=0,|j|=1/2$ and $n=0,|j|=3/2$. The case $|j|=1/2$ is valid for $g_\pm<1$ while the case $|j|=3/2$ is valid for $g_\pm<3$, as for larger values of the coupling constant the energies would become imaginary and an ultraviolet regularization would be needed to deal with the singularity of the Coulomb potential\cite{RC13,N07,PKC08}. By introducing this regularization, it can be seen that in the strong coupling regime (that in which the optical absorption energies of the excitons given by eq. (\ref{eq exc energies}) become imaginary) the excitonic bound energies become negative and the excitonic bound states ultimately merge with the continuum\cite{RC13,PKC08}.

\begin{figure*}
(a)
\begin{minipage}{.46\linewidth}
\includegraphics[scale=0.36]{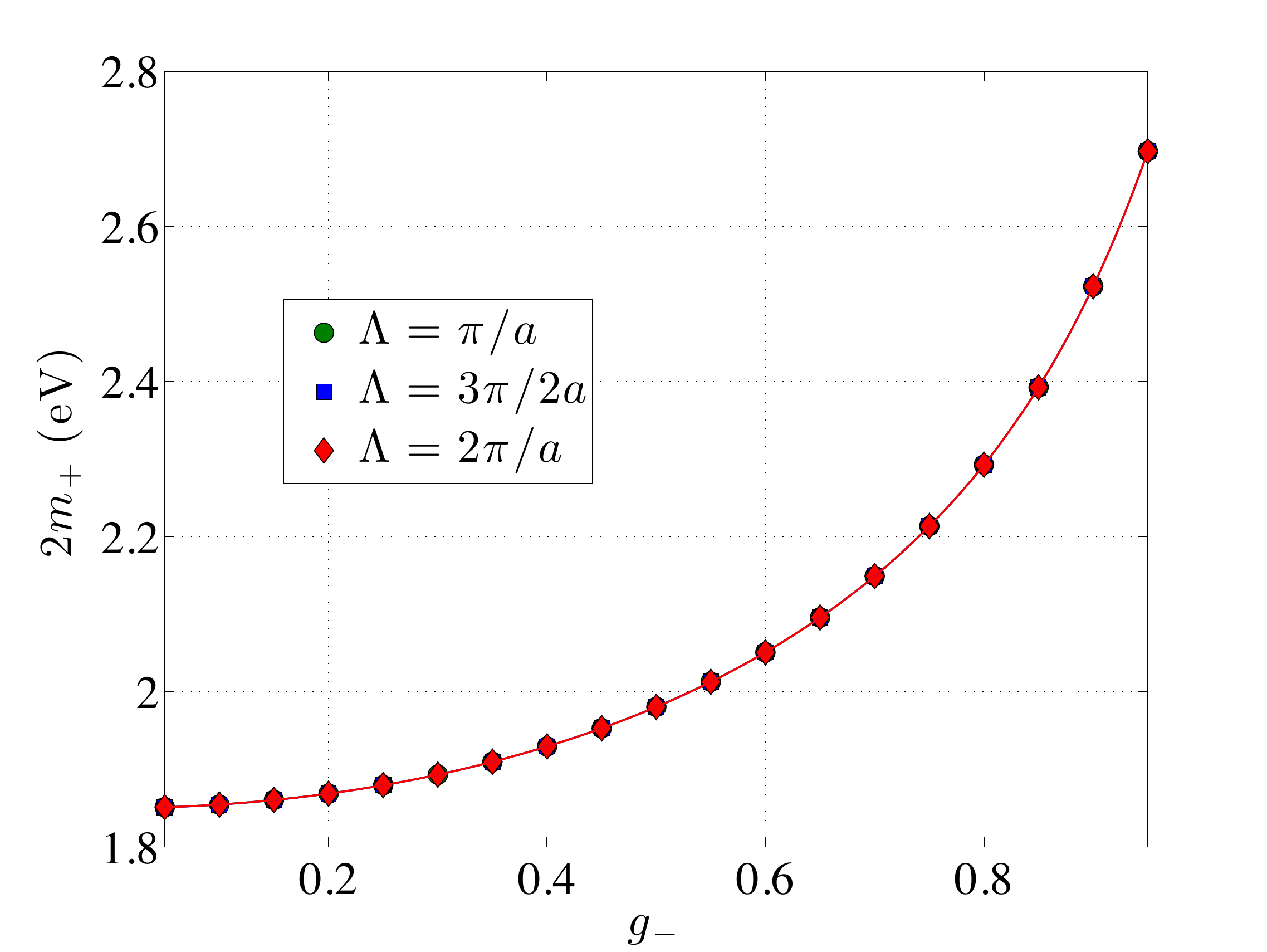}
\end{minipage}
(b)
\begin{minipage}{.46\linewidth}
\includegraphics[scale=0.36]{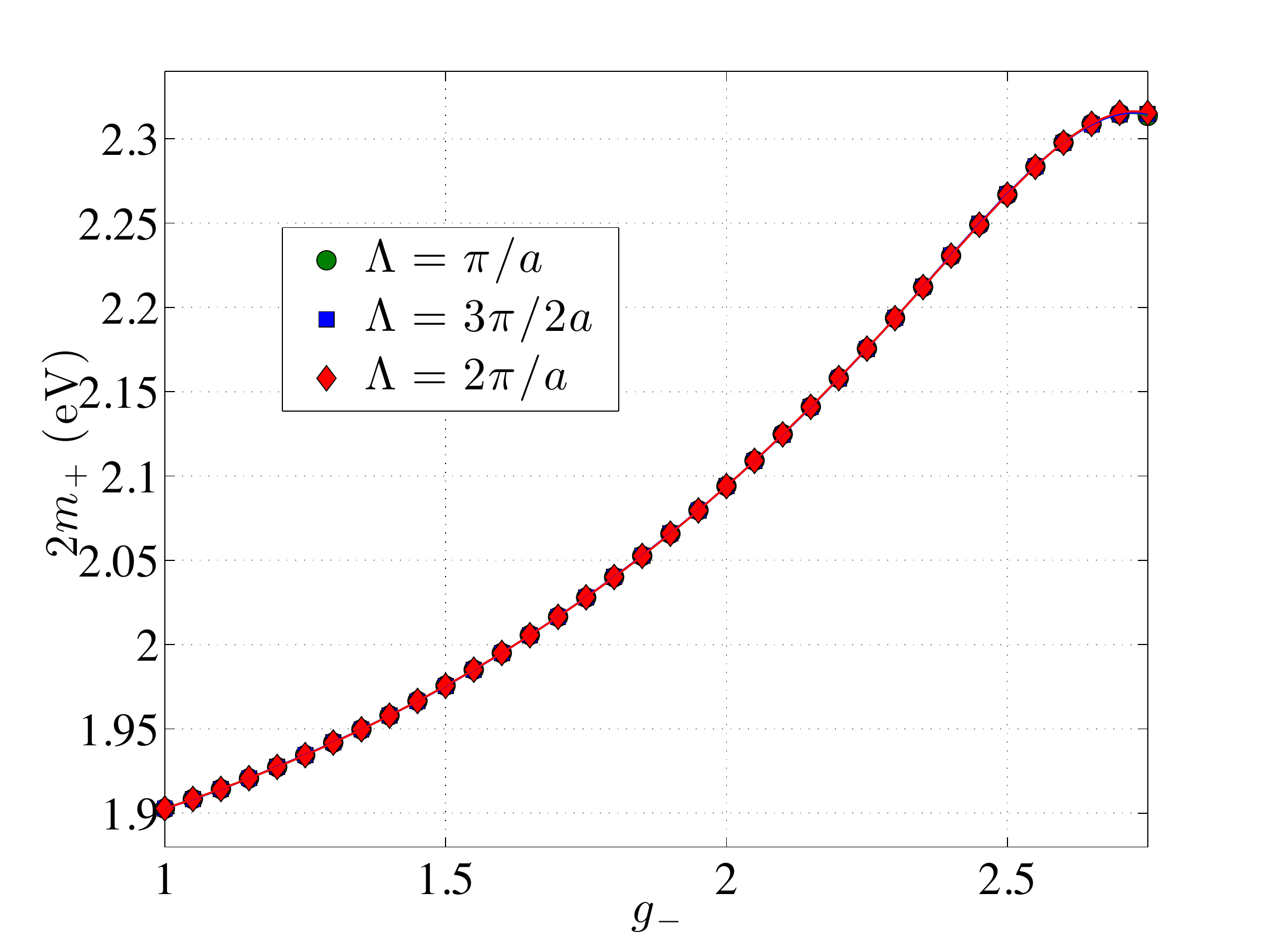}
\end{minipage}
(c)
\begin{minipage}{.46\linewidth}
\includegraphics[scale=0.36]{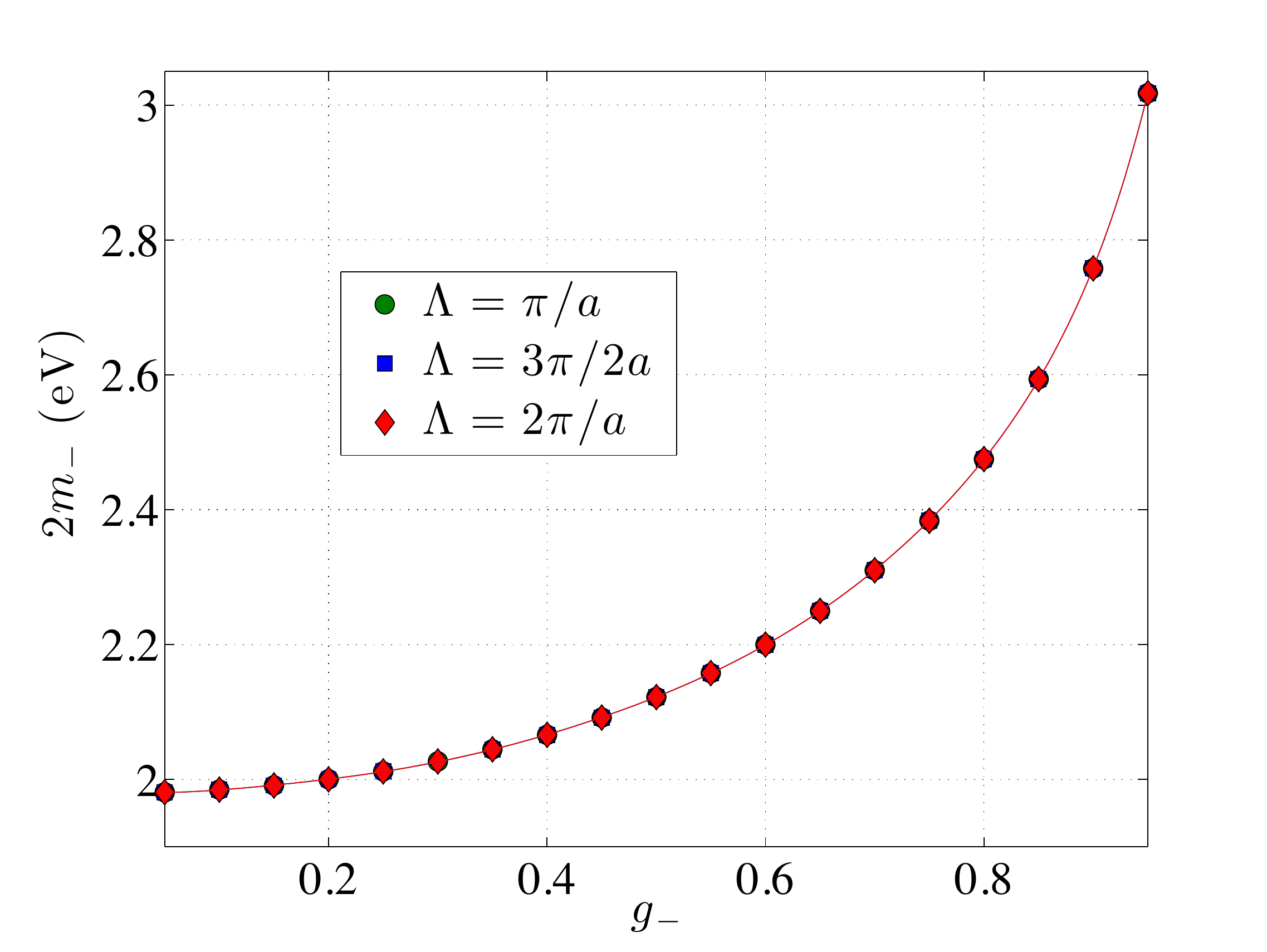}
\end{minipage}
(d)
\begin{minipage}{.46\linewidth}
\includegraphics[scale=0.36]{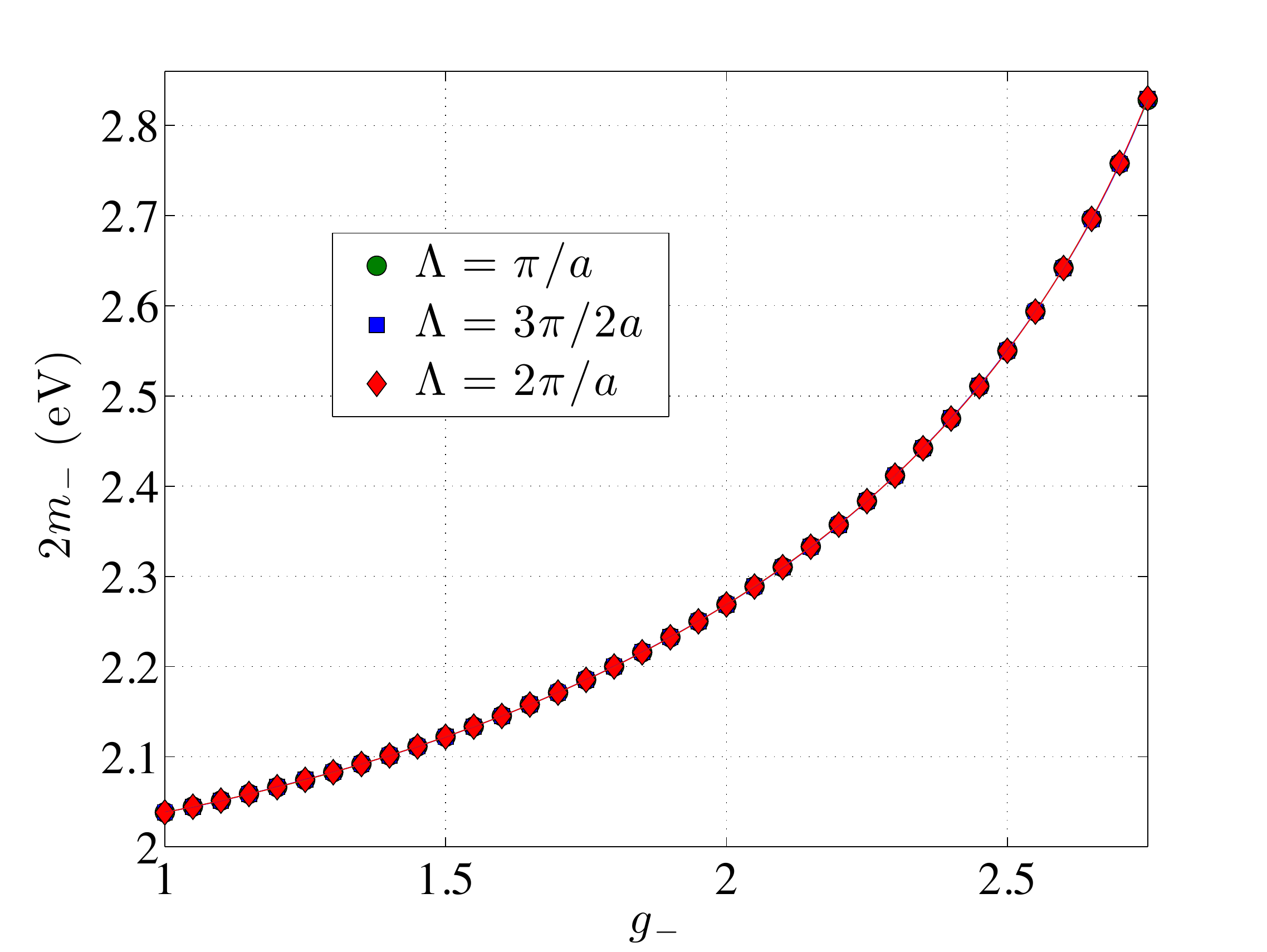}
\end{minipage}
\caption{Values for the $\tau=+1$ valley of the gap for the spin up electrons (a,b) and gap for the spin down electrons (c,d) as a function of the corrected coupling constant for spin down electrons $g_-$, for three different values of the momentum cut-off (in units of $\hbar$). }
\label{fig. 2}
\end{figure*}

This dramatic decrease of the excitonic energies in the strong coupling regime means that for $1<g_\pm<3$, the lowest excitonic state corresponding to the excitonic peaks A, B seeing in the experiments should no longer be $n=0,|j|=1/2$, but actually be the second lowest energy configuration $n=0,|j|=3/2$. Hence, we have an scenario in which for the small coupling regime, $g_\pm<1$, the observed peaks A, B correspond to the state $n=0,|j|=1/2$, while for the strong coupling regime $1>g_\pm<3$ they correspond to the state $n=0,|j|=3/2$:
\begin{equation}
E_{A,B}=E_{+,-}(n=0,|j|=1/2)\quad\quad g_\pm<1
\label{eq exp cond 1}
\end{equation}
\begin{equation}
E_{A,B}=E_{+,-}(n=0,|j|=3/2)\quad\quad 1<g_\pm<3
\label{eq exp cond 2}
\end{equation}
It has already been proposed in a previous work\cite{SK14} that the lowest bright excitons correspond to $|j|=3/2$ in the strong coupling regime $g_\pm>1$. In that work the ground state is claimed to be the lowest excitonic bound state $n=0,|j|=1/2$, and the transition from the ground state to the bright excitonic states is actually a transition between proper excitonic states. They concluded that only transitions to states with $|j|=3/2$ are allowed in the strong coupling regime.

Eqs. (\ref{eq exp cond 1},\ref{eq exp cond 2}) give two experimental conditions that have to be fulfilled by the renormalized parameters $\Delta,\lambda_{c,v}, v_\pm$. The dielectric permitivity $\epsilon$ that enters in the coupling constant $g_\pm$ can be written as $\epsilon=\epsilon_0(1+\epsilon_s)/2$, being $\epsilon_0$ the dielectric permittivity of the vacuum and $\epsilon_{s}$ the dielectric constant of the substrate. As the experimental absorption energies are obtained using a quartz substrate, we have $\epsilon_s=3.9$. Notice that Eq.(\ref{eq exc energies}) assumes equal hole and electron effective masses.

\begin{figure*}
(a)
\begin{minipage}{.46\linewidth}
\includegraphics[scale=0.36]{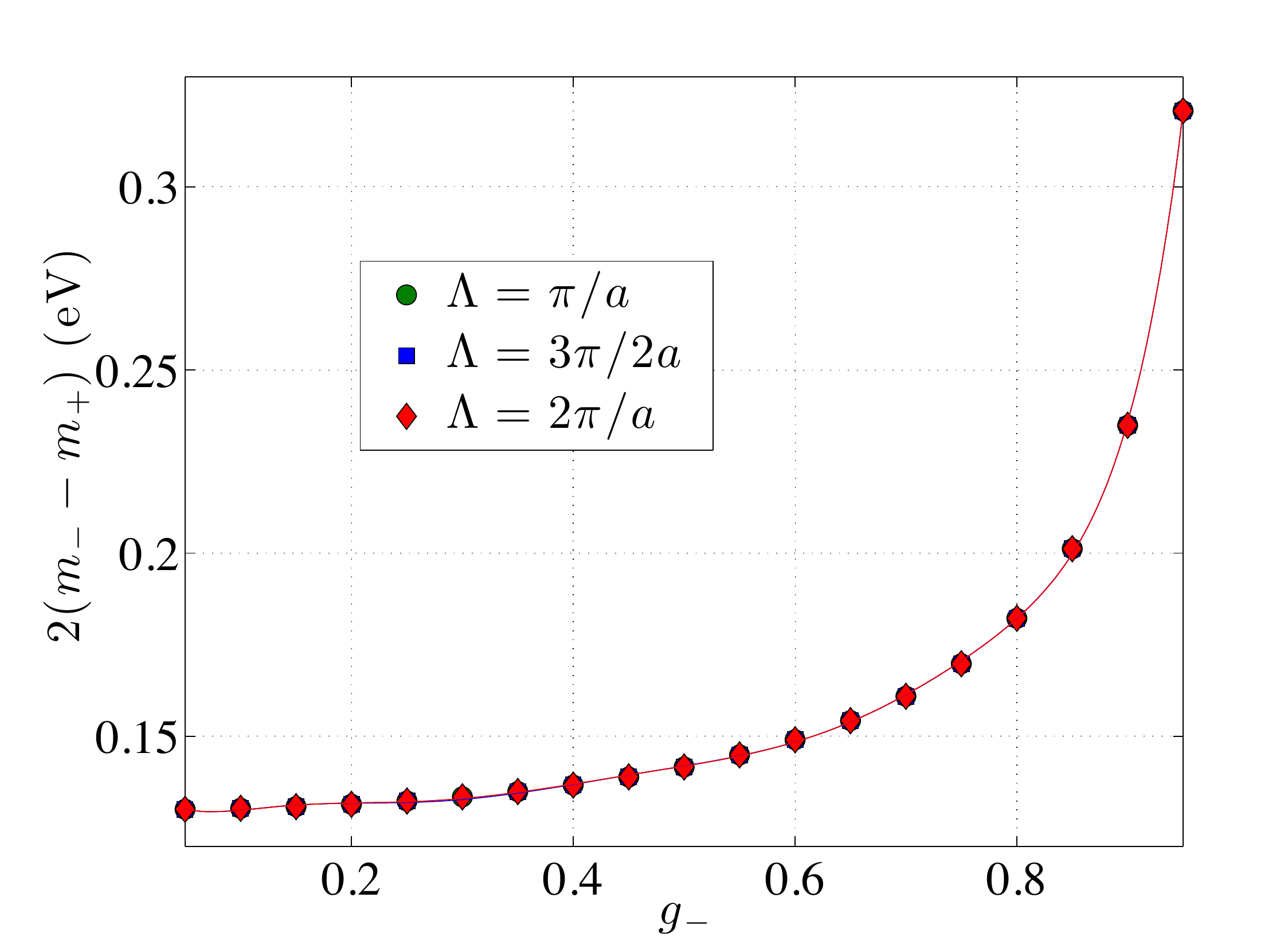}
\end{minipage}
(b)
\begin{minipage}{.46\linewidth}
\includegraphics[scale=0.36]{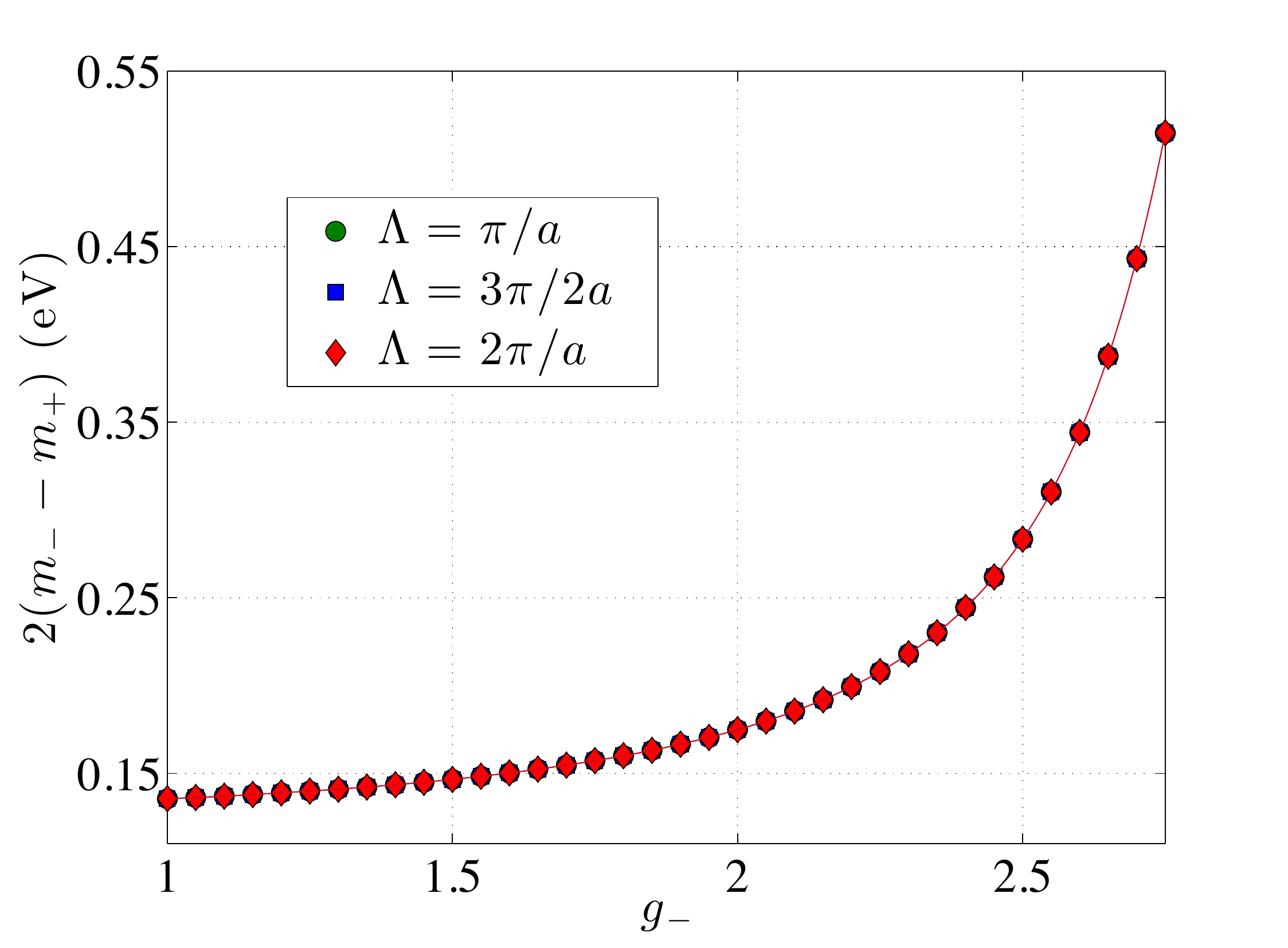}
\end{minipage}
(c)
\begin{minipage}{.46\linewidth}
\includegraphics[scale=0.36]{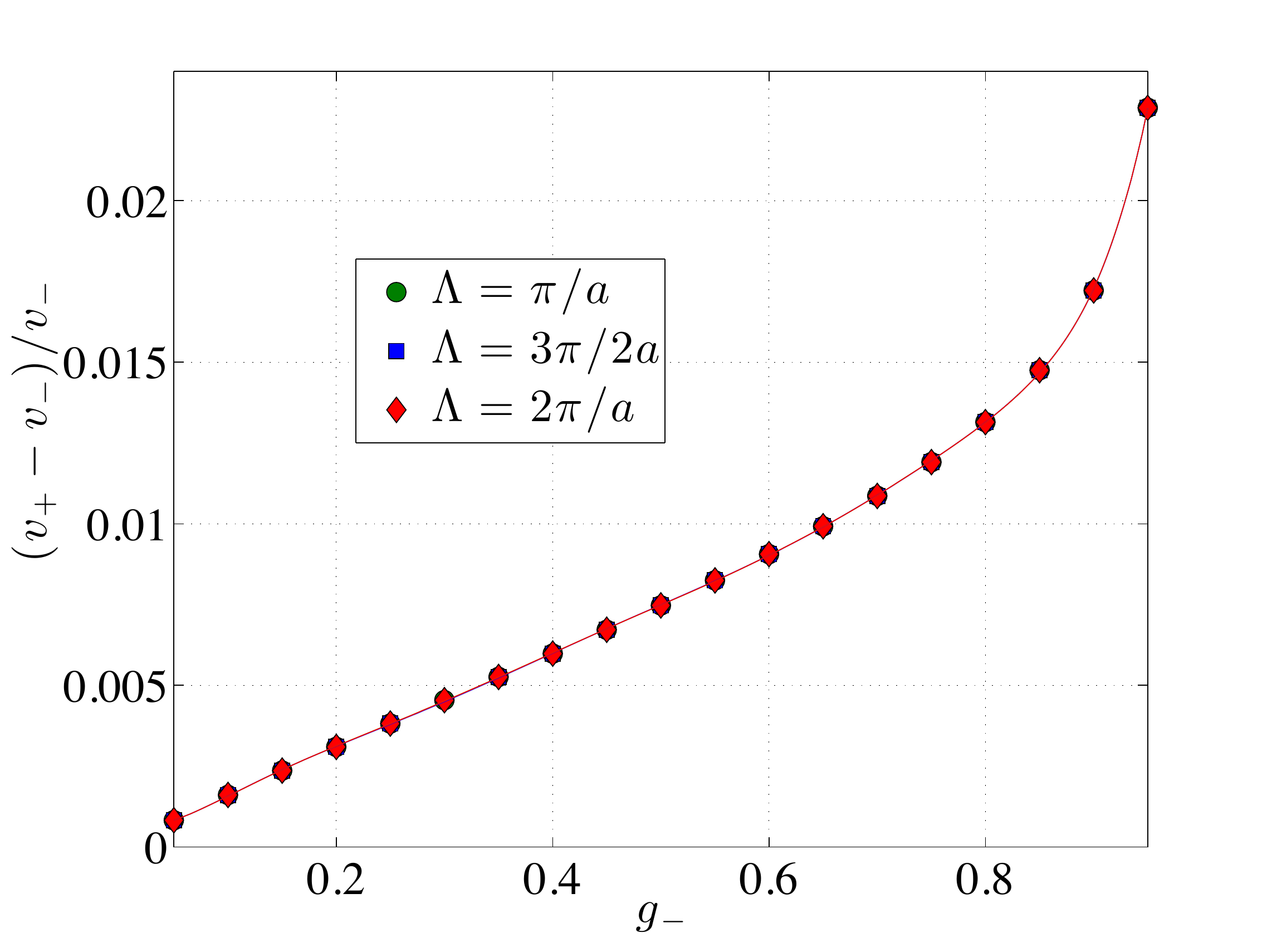}
\end{minipage}
(d)
\begin{minipage}{.46\linewidth}
\includegraphics[scale=0.36]{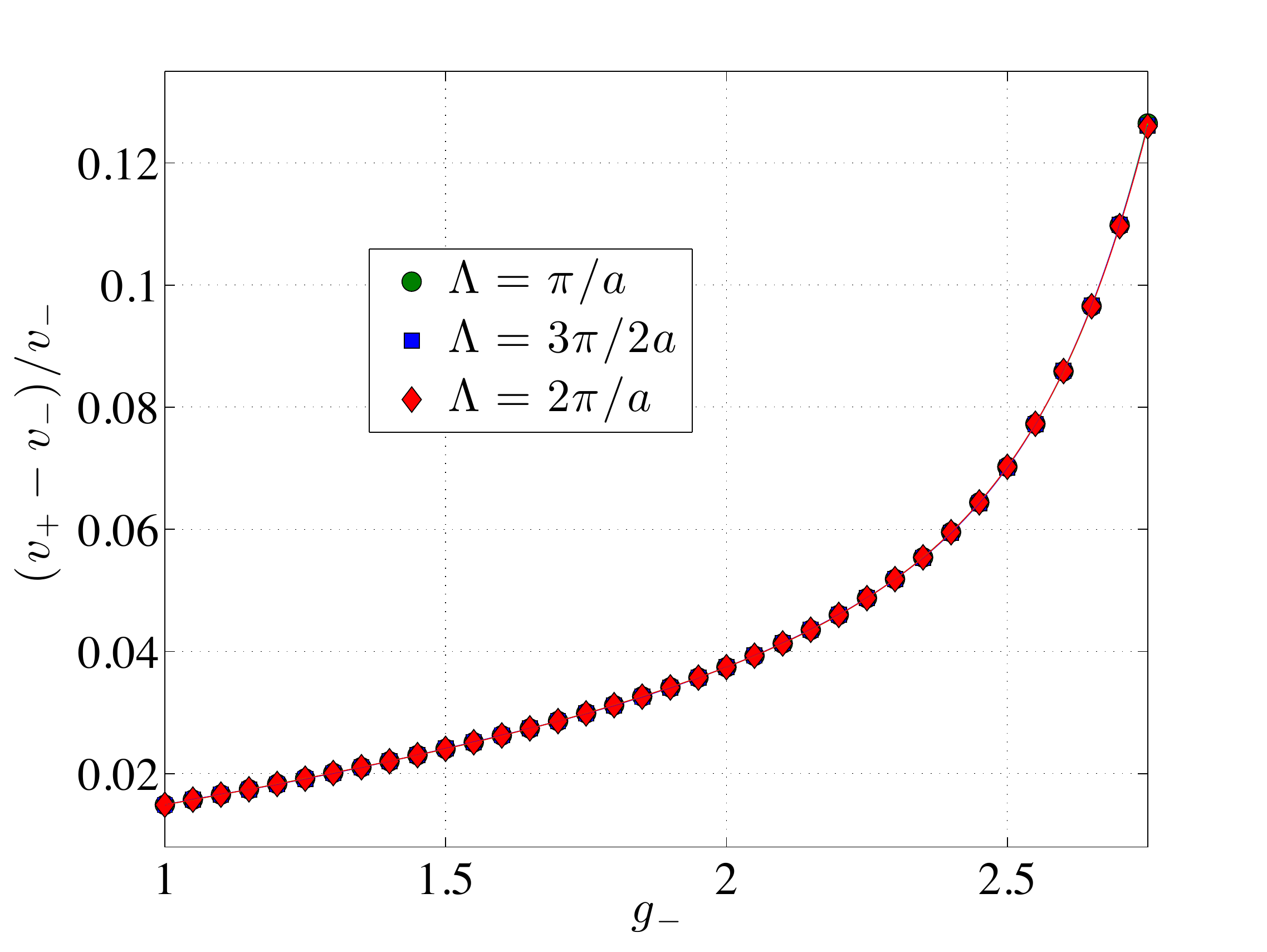}
\end{minipage}
\caption{Values for the $\tau=+1$ valley of the gap difference between spin down and up electrons (a,b) and normalized Fermi velocity difference between spin up and down electrons (c,d), as a function of the corrected coupling constant for spin down electrons $g_-$, for three different values of the momentum cut-off (in units of $\hbar$). }
\label{fig. 3}
\end{figure*}

As we already said, the energies in Eq. (\ref{eq exc energies}) are obtained by solving the two particle problem\cite{SSZ10}. This is a sensible approach since we are working in the instantaneous approximation, so the self consistent solution of the Schwinger-Dyson equations provides no term proportional to the frequency $p_{0}$ in Eq.(\ref{Ifunctions}). This implies that there is no wave function renormalization $Z_{\psi}$, and there is no loss of electronic coherence\cite{PFS13}. This fact, together with the absence of any imaginary part in $\Sigma_{\tau}(p)$, means that the poles in the full two-particle propagator will coincide with the ones in the non interacting two particle propagator, but dressed with the renormalized parameters in the Green functions\cite{GGG}.

Since the excitonic data are not enough to fully execute the renormalization program, we are forced to impose two additional conditions so that the renormalized parameters will depend only on one bare parameter and the cut-off. At very small interactions (at very high dielectric permittivity $\epsilon\rightarrow\infty$), we will force a zero conduction band splitting $\lambda_c=0$ and equal Fermi velocities for both spin projections $v_+=v_-$. The motivation for this choice comes from the fact that for $\epsilon\rightarrow\infty$ quantum corrections are negligible (assuming a finite physical cut-off), and the conduction band should remain (approximately) degenerate in spin due to the nature of the wave functions\cite{OR13}, while the hopping parameters should be insensitive to spin. In this limit the conditions on the renormalized parameters translate to conditions on the bare parameters: $\lambda_c^0=0$ and $v_{+}^0=v_{-}^0$.

Since we have imposed two extra conditions, we still have two free parameters: $v_{+}^{0}$/$v_{-}^{0}$ and the momentum cut-off $\Lambda$. We will solve the Schwinger-Dyson equations for different values of the cut-off and the Fermi velocity bare parameter, and give the results as a function of the renormalized coupling constant $g_-$.

\section{Solution of the Schwinger-Dyson equations}
\label{sec:solution}
We will present the results for one of the two valleys ($\tau=+1$). The values for the other valley are obtained by changing the sign of $\lambda_{c,v}$, and consequently interchanging the values of the masses ($m_+\leftrightarrow m_-$) and the Fermi velocities ($v_+\leftrightarrow v_-$).

In Figs.\ref{fig. 1},\ref{fig. 2},\ref{fig. 3} the values of different renormalized parameters as functions of $g_{-}$ are plotted for three different values of $\Lambda$, and for the two regimes $g_-<1$ and $1<g_-<3$. The general behavior when $g_{-}$ approaches to the critical values $g^{c}_{-}=1$ and $g^{c}_{+}=3$ will be modified with a proper regularization of the Coulomb potential. This regularization will make the calculations more involved but this treatment is possible within the presented theoretical framework. Also, this regularization allows for a more accurate computation of the critical value of $g_{-}$ corresponding to the point when the excitonic states will merge to the continuum, and the observed excitonic peaks are no longer expected to correspond to the lowest energy state $|j|=1/2$, but rather to the first excited state $|j|=3/2$. Also let us remember that all the plotted values are constrained to match the experimental energies of the excitonic peaks $E_A$ and $E_B$. As an example, for a coupling constant $g_-=2$ and a physical momentum cut-off $\Lambda=2\hbar\pi/a$ (the lattice constant is taken to be $a=3.193$ \r{A}\cite{XLF12}), we have an scenario with a conduction band splitting $2|\lambda_c|\approx75$ meV, fully compatible with the measured optical absorption energies of the excitonic peaks A, B.


Both the gaps and the difference between spin up and down Fermi velocities are cut-off independent. For the Fermi velocity difference, the independence on the cut-off $\Lambda$ can be derived from eqs. (\ref{eq. Fermi velocity},\ref{SDeqs2}). The correction to the Fermi velocity is given by $\delta v_s=v_sI_s$, and if we do $\delta v_+-\delta v_-$, the two logarithms which give the explicit dependence on $\Lambda$ are subtracted, so the explicit dependence vanishes. There is however an implicit dependence on $\Lambda$ coming from $v_s$ and $m_s$. From conditions of eqs.  (\ref{eq exp cond 1},\ref{eq exp cond 2}) we obtain $m_s$ as a function of $v_s$, so all the implicit dependence on the cut-off lies in $v_s$. With the extra condition $v_{+}^{0}=v_{-}^{0}$, we can write $v_+=v_-+\delta v_+-\delta v_-$, and for any given $v_-$ we have an equation for $\delta v_+-\delta v_-$ independent of $\Lambda$. With the Fermi velocity difference being cut-off independent, from conditions of eqs.  (\ref{eq exp cond 1},\ref{eq exp cond 2}) one automatically obtains cut-off independent masses for each spin.

\section{Summary}
\label{sec:summary}
Large conduction band and Fermi velocity spin splittings are found due to the effect of Coulomb interactions, fully consistent with optical absorption measurements. The ultimate reason of these splittings is the presence of a different gap for the two spin polarization species, product of spin-orbit interaction, which induces a different renormalization of the gap and Fermi velocities of the two spin projections. To ensure consistency with absorption experiments, we used the values of the measured energies of the excitonic peaks of MoS$_{2}$ on a quartz substrate as renormalization conditions.


\section{Acknowledgments}
A. C. gratefully acknowledges conversations with  M. A. H. Vozmediano and R. Asgari at the early stages of this project, and with E. Cappelluti. This research is partially supported by CSIC JAE-doc fellowship program and the Spanish MECD Grants No. FIS2011-23713 and No. PIB2010BZ-00512.

\appendix
\section{Computation of the fermionic self-energy}
\label{app}

To obtain the self-energy we need to compute the integral of eq. (\ref{integral}):
\begin{equation}
\Sigma(p)\equiv\left(\begin{array}{cc}I_+(p)&0\\0&I_-(p)\end{array}\right)=e^2\int\frac{d^3q}{(2\pi)^3}D(q)G_{+1}(p-q),
\label{integral app}
\end{equation}
with, keeping terms up to first order in $p$:
\begin{equation}
I_s(p)=-\left(\begin{array}{cc}I^{(1)}_s&0\\0&I^{(2)}_s\end{array}\right)-v_s I_{s}(\sigma_xp_x+\sigma_yp_y).
\end{equation}

First we shall obtain the asymptotic behavior of the scalar field self-energy $\Pi(q)$, given by eq. (\ref{polarizationfunction}), in the limits of low and high momenta. We work in the instantaneous approximation $q_0=0$. For $v_s|\vec{q}|<<m_s$ we have:
\begin{equation}
\Pi(|\vec{q}|)=\frac{e^2}{3\pi m_r}|\vec{q}|^2,
\label{eq. exp. 1}
\end{equation}
while in the opposite limit, $v_s|\vec{q}|>>m_s$:
\begin{equation}
\Pi(|\vec{q}|)=\frac{e^2}{4v_r}|\vec{q}|
\label{eq. exp. 2}
\end{equation}

Now, from the low energy Hamiltonian (\ref{initialHam}) one can extract the inverse bare fermion propagator. Inverting it and replacing the bare parameters by the renormalized ones, we get the full fermion propagator which enters the integral (\ref{integral app}). Inserting the explicit values of $D(q)$ and $G_{+1}(p-q)$ we have:
\begin{widetext}
\begin{eqnarray}
I^{(j)}_s&=&i2\pi g_s\left(\int_0^{m_s}\frac{d|\vec{q}|}{2\pi}\big(1+\frac{4g_s}{6m_r}|\vec{q}|\big)^{-1}+(1+\pi g_r/2)^{-1}\int_{m_s}^{\Lambda v_s}\frac{d|\vec{q}|}{2\pi}\right)\nonumber\times\\
&\times &\int_{-\infty}^\infty\frac{dq_0}{2\pi}g_0^s(q)\left(q_0+i(-1)^j\frac{\Delta}{2}+s\,i\Big((j-1)\lambda_c+(2-j)\lambda_v\Big)\right),\\
I_s&=&2\pi g_s\left(\int_0^{m_s}\frac{d|\vec{q}|}{2\pi}\big(1+\frac{4g_s}{6m_r}|\vec{q}|\big)^{-1}+(1+\pi g_r/2)^{-1}\int_{m_s}^{\Lambda v_s}\frac{d|\vec{q}|}{2\pi}\right)\int_{-\infty}^\infty\frac{dq_0}{2\pi}g_0^s(q)\left(1-|\vec{q}|^2g_0^s(q)\right),
\end{eqnarray}
\end{widetext}
with:
\begin{equation}
g_0^s(q)=\left(\Big(q_0+s\,\frac{i}{2}(\lambda_c+\lambda_v)\Big)^2+|\vec{q}|^2+m_s^2\right)^{-1}.
\end{equation}
Note that we divided the momentum integral in two regions, one for $v_s|\vec{q}|<<m_s$ and another for $v_s|\vec{q}|>>m_s$, using the expansions of eqs. (\ref{eq. exp. 1}) and (\ref{eq. exp. 2}). Note also that we introduced a momentum cut-off $\Lambda$, as the momentum integrals diverge. When doing the integrals in $q_0$ one should be careful and expand the results in $|\vec{q}|$ accordingly for each momentum integral region $v_s|\vec{q}|<<m_s$ and $v_s|\vec{q}|>>m_s$. Doing the integrals we finally arrive to:

\begin{subequations}
\begin{equation}
I^{(j)}_s=(-1)^{j+1}I^{z}_s,
\end{equation}
\begin{equation}
I^{z}_s=\frac{3m_r}{4}\ln\left(1+\frac{2g_sm_s}{3m_r}\right)+\frac{g_sm_s}{2+\pi g_r}\ln\left(\frac{\Lambda v_s}{m_s}\right),
\end{equation}
\begin{equation}
I_s=\frac{3m_r}{4m_s}\ln\left(1+\frac{2g_sm_s}{3m_r}\right)+\frac{g_s}{4+2\pi g_r}\ln\left(\frac{\Lambda v_s}{m_s}\right),
\end{equation}
\end{subequations}
which are the expressions used in the main text.


\end{document}